# Smart Farming Using IoT for efficient crop growth (2023)

Dr.A.Bharathi Sankar, Professor, Department of ECE, VIT Chennai(email: bharathsankar.a@vit.ac.in)

I.M.kawsalyaa, student, ECE, VIT Chennai(email: imkawsalyaa@gmail.com)

*Abstract*— **In general, automated crop irrigation systems make decisions based on static models built from the properties of the plant. In contrast, irrigation decisions in our suggested method are dynamically changed based on changing environmental conditions. The model's learning process reveals the mathematical links between the environmental factors employed in determining the irrigation habit and gradually improves its learning technique as irrigation data accumulates in the model.To analyse overall system performance, we constructed a test environment for the sensor edge, mobile client, and decision service in the cloud.** *Index Terms— automated, irrigation, sensor edge*

## I. INTRODUCTION

Smart farming encourages farming practises and methods that help farmers and resources last. It is economically practical, preserves soil quality, prevents soil degradation, conserves water, increases land biodiversity, and assures a natural and healthy environment.

Regrettably, not all regions of the Earth's surface are appropriate for agriculture due to a variety of constraints such as soil quality, topography, temperature, climate, and the fact that most significant cultivable areas are not homogeneous. Furthermore, each agricultural field has various important properties, such as soil type, irrigation flow, nutrient presence, and insect resistance, which are all measured independently in terms of quality and quantity for a specific crop. Crop rotation and an annual crop growth development cycle require both spatial and temporal changes to optimise crop yield in the same land.

Farmers in traditional farming practises visit their fields regularly throughout the crop's life in normal farming chores to better understand crop conditions. Current sensor and communication technologies provide a detailed view of the field, allowing farmers to identify ongoing field activities without physically being in the field. Wireless sensors monitor crops with more accuracy and spot problems at an earlier stage, often easing the use of smart instruments from crop sowing to harvest.

The timely application of sensors has made the entire farming process smart and cost-effective.

Because of careful supervision, it is effective. Sensors are fitted to the various autonomous harvesters, robotic weeders, and drones to collect data at regular intervals. However, the immensity of agriculture places high demands on technical solutions for long-term sustainability with minimal environmental impact. Sensor technology, combined with wireless connection, enables farmers to learn about the many demands and requirements of crops without having to leave their fields, allowing them to take remote action.

## II. EXISTING SYSTEM

The existing system begins to monitor the humidity and moisture levels. The sensors detect the level of water, and if it falls below a certain threshold, the system begins watering automatically. The sensor performs its function based on the change in temperature level. IoT also displays humidity and moisture level information, as well as date and time. Temperature levels can also be modified dependent on the sort of crops grown.

## III. PROPOSED SYSTEM

Flow, pressure, temperature, humidity, LDR, and soilmoisture sensors are used in our suggested system. The flow sensor was employed here to control the flow and amount of water fed to the plant. The pressure sensor is used to measure the weight of the soil, i.e. the force felt in the soil, which varies depending on the moisture content of the soil. The DHT11 sensor is used to measure the temperature and humidity of the atmosphere. Soil moisture is measured with a soil moisture sensor. The Rain sensor detects whether or not precipitation has occurred at that moment. If precipitation is detected, the water pump is turned off because it is not essential for it to be turned on. The data was shown on an LCD to the plant handlers.

### A. BLOCK DIAGRAM

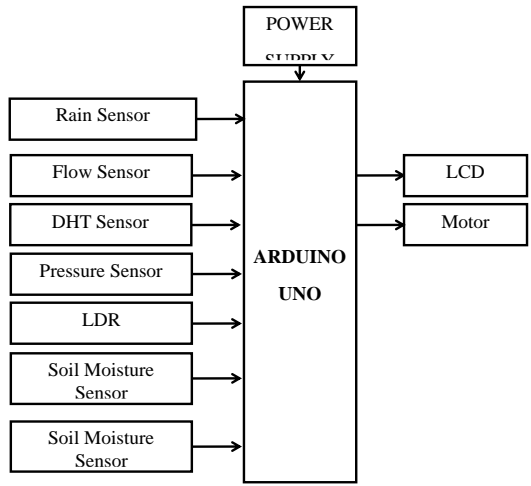

### B. MATERIALS REQUIREMENT

- Flow sensor
- DHT11 Sensor
- Pressure Sensor
- LDR
- Soil Moisture Sensor
- LCD
- Motor
- Power Supply
- Arduino UNO
- Arduino IDE

## IV. SENSORS USED IN THIS PROJECT

### A. dht11

- Sensor for humidity and temperature DHT11. The DHT11 is a basic digital temperature and humidity sensor that is inexpensive.
- The DHT11 is a single-wire digital humidity and temperature sensor that sends humidity and temperature information serially over a single wire.
- The DHT11 sensor measures relative humidity in percentage (from 20% to 90% RH) and temperature in degrees Celsius (0 to 50 °C).
- The DHT11 sensor employs a resistive humidity measurement component as well as an NTC temperature measurement component.

DHT11 Connector Pinout:

- DHT11 is a 4-pin sensor, these pins are VCC, DATA, GND and one pin is not in use shown in fig below.

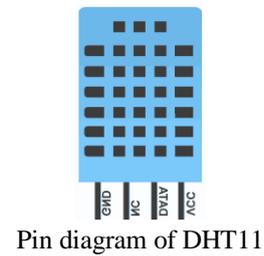

Pin diagram of DHT11

PIN DESCRIPTION:

|   | PIN NAME | PIN DESCRIPTION |
|---|----------|-----------------|
| 1 | VCC | Power supply 3.3 to 5.5 Volt DC |
| 2 | DATA | Digital output pin |
| 3 | NC | Not in use |
| 4 | GND | Ground |

COMMUNICATION WITH MICROCONTROLLER COMMUNICATION WITH MICROCONTROLLER

- DHT11 uses only one wire for communication. The voltage levels with certain time value defines the logic one or logic zero on this pin.
- The communication process is divided in three steps, first is to send request to DHT11 sensor then sensor will send response pulse and then it starts sending data of total 40 bits to the microcontroller.

START PULSE (REQUEST)

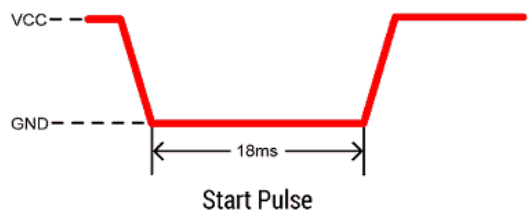

• To give a start pulse, pull down (low) the data pin for at least 18ms before pulling up, as illustrated in the diagram.

• To begin communication with DHT11, we must first deliver the start pulse to the DHT11 sensor.

RESPONSE

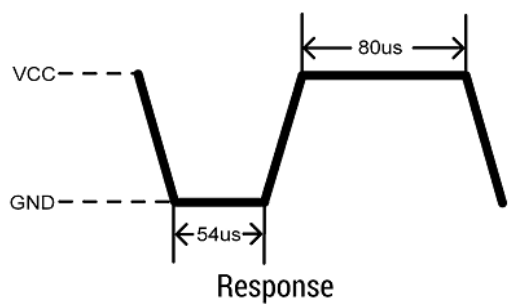

• After receiving the start pulse, the DHT11 sensor transmits a return pulse indicating that the DHT11 received the start pulse.

• The reaction pulse is low for 54us before rising to 80us.

DATA

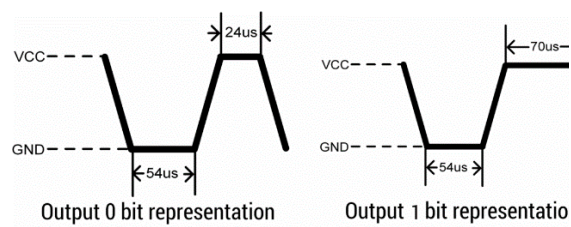

- • Following the response pulse, the DHT11 sensor provides data, which includes humidity and temperature values as well as a checksum.
- The data frame is 40 bits long in total, with 5 segments (bytes) each 8 bits long.
- The first two segments of these five segments contain humidity values in decimal integer form. This number represents Relative Percentage Humidity. The first eight bits are integers, and the next eight bits are fractions.
- The next two parts include the temperature in decimal integer form. This value represents the temperature in Celsius.
- The last segment is the checksum, which contains the checksums of the preceding four segments.
- In this case, the checksum byte is the direct summation of the humidity and temperature values. And we may check whether it matches the checksum value or not. If it is not equal, there is an error in the data received.
- When data is received, the DHT11 pin enters low power consumption mode until the next start pulse.

THE END OF THE FRAME

- After providing 40 bits of data, the DHT11 sensor sends a low level of 54us before going high. DHT11 enters sleep mode after this.

DHT11 versus DHT22

- Two DHT sensor variations that look identical and have the same pinout but have different characteristics and specifications:
- DHT11 • Extremely low cost
- 3 to 5V power and I/O
- 2.5mA max current use during conversion (while requesting data)
- Good for 20-80% humidity readings with 5% accuracy
- Good for 0-50°C temperature readings ±2°C accuracy

• A sampling rate of no more than 1 Hz (once per second)
• 15.5mm x 12mm x 5.5mm body size • 4 pins with 0.1" spacing

B. raindrop detector
Raindrop Sensor is a gadget for detecting rain. It is made up of two modules: a rain board that detects rain and a control module that compares and converts the analogue value to a digital value. Raindrop sensors are used in the automotive industry to automatically regulate windscreen wipers, in agriculture to detect rain and in home automation systems.

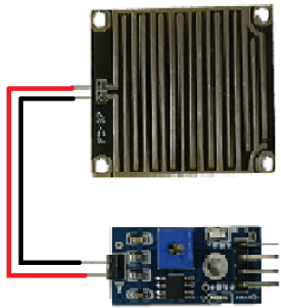

The raindrop sensor's control module has four outputs. VCC is linked to a 5V power supply. The module's GND pin is linked to ground. The D0 pin is connected to the microcontroller's digital pin for digital output, or the analogue pin can be used. To use the analogue output, connect the A0 pin to the ADC pin of a microcontroller. Because Arduino has six ADC pins, we can use any of them directly without utilising an ADC converter. A potentiometer, an LN393 comparator, LEDs, capacitors, and resistors make up the sensor module. The components of the control module are depicted in the pinout image above. The rain board module is made up of copper tracks that function as variable resistors. Its resistance changes with the amount of moisture on the rain board. The rain board module is depicted in the figure below..

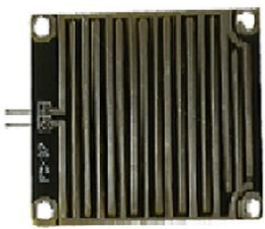

**Rain Board Module**

*c. water flow sensor*
yfs201 hall effect water flow sensor with arduino for detecting water flow rate and volume. this is a fantastic project that may be utilised in industry, at home, or in water flow measurement application in a water tap, tunnel, river, and so on. the circuit diagram for the water flow sensor for flow rate & volume measurement using arduino code is shown below.

this project can be used to measure liquid flowing via a pipe or container, as well as to design a control system based on the rate or quantity of water flow. for example, you might use this to quantify the amount of water used to water your plants in order to save waste. you may also use it to create water dispenser machines used in industry and beverages.

check out the advanced version of this project to remotely monitor water flow rate and volume: iot esp8266-powered water flow metre

*WATER FLOW SENSOR FOR FLOW RATE & VOLUME MEASUREMENT USING ARDUINO:*
BILL OF MATERIALS

| S.N. | Components Name | Quantity | Purchase Links |
|---|---|---|---|
| 1 | Arduino UNO Board | 1 | Amazon \| AliExpress |
| 2 | YFS201 Hall Effect Water Flow Sensor | 1 | Amazon \| AliExpress |
| 3 | 16x2 LCD Display | 1 | Amazon \| AliExpress |
| 4 | Potentiometer 10K | 1 | Amazon \| AliExpress |
| 5 | Connecting Wires | 20 | Amazon \| AliExpress |
| 6 | Breadboard | 1 | Amazon \| AliExpress |
| | | | |

*CIRCUIT DIAGRAM:*

Connect LCD pins 1, 3, 5, 16, and 17 to GND and 2, 15 to 5V VCC. Connect the LCD pins 4, 6, 11, 12, 13, 14, to the Arduino digital pins D7, D6, D5, D4, D3, D2.

Connect the VCC pins of the YFS201 HALL EFFECT WATER FLOW SENSOR to a 5V power supply and the GND pins to GND. Because it is an analogue sensor, connect its analogue pin to Arduino A0, as indicated in the diagram above.

a) YFS201 WATER FLOW SENSOR WITH HALL EFFECT:INTRODUCTION:

This sensor connects to your water pipe and contains a pinwheel sensor that measures the amount of liquid that has passed through it. With each revolution, an embedded magnetic hall effect sensor generates an electrical pulse. The hall effect sensor is isolated from the water pipe, allowing it to remain safe and dry.

Check out the datasheet for the YF-S201 Hall Effect Water Flow Metre / Sensor.

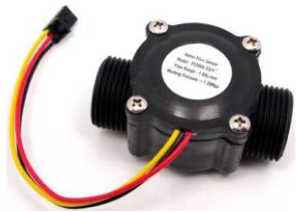

- There are three wires on the sensor: red (5-24VDC power), black (ground), and yellow (Hall effect pulse output). The WATER FLOW RATE can be simply calculated by counting the pulses from the sensor's output. Each pulse contains about 2.25 millilitres. It should be noted that this is not a precision sensor, and the pulse rate varies slightly based on the flow rate, fluid pressure, and sensor orientation. If more than 10% precision is desired, it will require rigorous calibration. It is, nonetheless, excellent for basic measurement chores!

D. Pressure sensor

- This HX710B pressure sensor module employs a high-precision AD sampling chip, an 0-40KPa air pressure sensor, a 2.5mm hose connector, and the ability to detect water level and other air pressure. This air pressure module employs a high-precision AD sampling chip, an 0-40KPa air pressure sensor, a 2.5mm hose connector, the ability to detect water level and other air pressure, a voltage range of 3.3V-5V, a compact design, and the usage of a 5k resistance bridge. The sensor, as well as the specific pressure value, must be computed by ourselves.Features/Specs:
- • Model: HX710B • Interface Type: Serial • Data Output Rate: 10 SPS/80 SPS • Operating Current: 1.5mA • Operating Temperature Range: -40 to 85°C
- Two distinct differential input channels.
- Active Low Noise PGA on-chip with gain selections of 32, 64, and 128.
- Load Cell and ADC Analogue Power Supply On-Chip Power Supply Regulator.
- Power-On-Chip Reset
- Simple Digital Control And Serial Interface.
- IC Package: SOP 8
- Length: 18mm
- Width: 17mm
- Height: 2mm
- Weight: 2gm

*E. ldr*
As the name implies, an LDR (Light Dependent Resistor) is a type of resistor that operates on the photoconductivity principle, which means that resistance varies with the intensity of light. Its resistance reduces as the intensity of the light increases. It is frequently used as a light sensor, light metre, automatic street light, and in other applications where light sensitivity is required. It is also known as a Light Sensor.Light-dependent resistors are built of photosensitive semiconductor materials such as Cadmium Sulphides (CdS), lead sulphide, lead selenide, indium antimonide, or cadmium selenide and are arranged in a Zig-Zag pattern, as seen in the image below. Two metal contacts are added on opposite ends of the Zig-Zag form to aid in the connection with the LDRs.

Now, a transparent coating is placed to the top to cover the zig-zag-shaped photosensitive material, and because the coating is transparent, the LDR may capture light from the outside environment for operation.

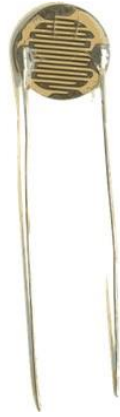

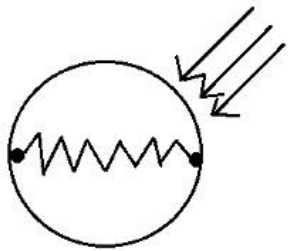

LDR Symbol

WORKING PRINCIPLE OF LDR

When light falls on its photoconductive material, it absorbs its energy, and the electrons in the valence band of that photoconductive material become excited and move to the conduction band, increasing the conductivity as the light intensity increases.

Furthermore, the incident light energy must be larger than the bandgap gap energy in order for the electrons in the valence band became energised and moved to the conduction band.

The LDR has the largest resistance in the dark, around 1012 Ohm, and this resistance diminishes as the light level increases.RESISTANCE V/S LIGHT INTENSITY

According to LDR properties, the amount of light entering the LDR is inversely proportional to the sensor's resistance, and the graph is hyperbolic in form.

TYPES OF LDR OR PHOTORESISTORS

1. INTRINSIC PHOTORESISTOR

This photoresistor is built of pure semiconductors with no doping. Pure semiconductors such as silicon and germanium are used in this type of photoresistor. When incoming light with sufficient energy strikes this, electrons gain that energy and become excited, and a few of them move to the conduction band.2. PHOTORESISTOR EXTRINSIC

This form of photoresistor employs a doped semiconductor, which implies that certain impurities, such as phosphorus, are mixed with the semiconductor to create the photoresistor.

Extrinsic light-dependent resistors are often constructed for longer light wavelengths, with a preference for infrared (IR).

A digital output, such as an LED or a relay, is controlled by this device.

The spectral sensitivity of an LDR relates to the range of light wavelengths that the LDR can detect.

sensitive to. LDRs have varying spectrum sensitivity, with some being more sensitive to visible light and others being more sensitive to infrared light.

LDRs are often temperature sensitive, and their resistance varies dramatically with temperature. As a result, it is critical to select an LDR that is rated for the temperature range of the application.

LDRs are not normally waterproof and can be damaged by damp or water. As a result, they should be utilised in dry locations or enclosed in a waterproof enclosure.

Yes, by translating the resistance of the LDR to a voltage using a simple voltage divider circuit, LDRs may be used to measure light intensity. The output voltage can then be measured.

measured with a voltmeter or digitally measured with an analog-to-digital converter (ADC).

Because they are only sensitive to the intensity of light and not the colour, light-dependent resistors are not commonly employed for colour sensing. Colour sensors often employ more complicated electronics that are sensitive to specific wavelengths of light, such as photodiodes or phototransistors.

LDRs are not commonly utilised for motion detection since they only respond to changes in light intensity rather than movement. Other technologies, such as passive infrared (PIR) or ultrasonic sensors, are commonly used in motion sensors.

Soil moisture semsor F

When it comes to designing your smart irrigation system or autonomous plant watering system, the SOIL MOISTURE SENSOR is the first item that comes to mind. With this sensor and a little Arduino help, we can create a system that will water your plants only when necessary, preventing overwatering and underwatering.

In this post, we will INTERFACE THE SOIL MOISTURE SENSOR WITH AN ARDUINO and determine the volumetric concentration of water within the soil. This sensor is configured to output data in both digital and analogue modes. We will read this data and use an LED to display the output status for digital output and a serial monitor or an LED with PWM for analogue output.

.

> https://orcid.org/0009-0005-3704-5266<

*SOIL MOISTURE SENSOR PINOUT*

VCC, GND, Aout, and Dout are the four pins on the soil moisture sensor. These four pins can be utilised to obtain soil moisture data from the sensor. The SOIL MOISTURE SENSOR PINOUT is as follows:

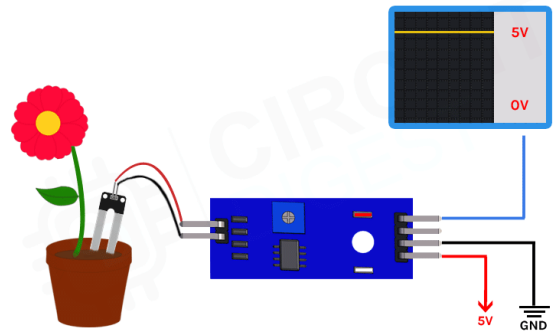

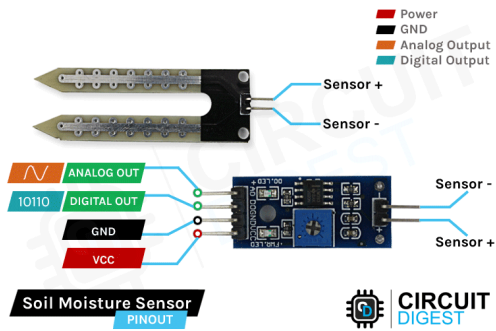

VCC is the soil moisture sensor's power supply pin, which can be linked to either 3.3V or 5V of the supply. However, keep in mind that the analogue output will change depending on the supplied input voltage.

GND is the board's ground pin, and it should be connected to the Arduino's ground pin.

DOUT is the board's DIGITAL OUTPUT PIN; low indicates suitable soil moisture, and high indicates low soil moisture.

AOUT is the board's ANALOGUE OUTPUT PIN, which will provide us with an analogue signal between vcc and ground.

.

The GIF ANIMATION OF SOIL MOISTURE SENSOR shown above shows how the analogue output of the sensor varies when the water level in the soil changes. When water is introduced to the soil, the voltage reduces from 5V to 0V. When water is put to the soil, the signal LED on the board illuminates. To keep things simple, we haven't showed how the digital pin works in the GIF above. When water is applied to the soil, the digital pin changes from low (0V) to high (5V) using the on-board G. Uno Arduino

The Arduino Uno (datasheet) is a microcontroller board based on the ATmega328. It contains 14 digital I/O pins (six of which are PWM outputs), 6 analogue inputs, a 16 MHz ceramic resonator, a USB connection, a power jack, an ICSP header, and a reset button. It comes with everything you need to support the microcontroller; simply connect it to a computer through USB or power it using an AC-to-DC adapter or battery to get started. The FTDI USB-to-serial driver chip is absent from the Uno, as is the case with all previous boards. It instead employs an Atmega16U2 (Atmega8U2 up to version R2) programmed as a USB-to-serial converter. The Uno board revision 2 includes a resistor that pulls the 8U2 HWB line to ground, making it easier to enter DFU mode. The board's third revision includes the following new features:

• 1.0 pinout: introduced SDA and SCL pins near the AREF pin, as well as two other new pins near the RESET pin, the IOREF, that allow the shields to adapt to the voltage supplied by the board. Shields will be compatible in the future with both the AVR-based boards that operate at 5V and the Arduino Due, which operates at 3.3V. The second is an unconnected pin that will be used in the future.
• A more powerful RESET circuit.
• The 8U2 is replaced by an Atmega 16U2. "Uno" means "one" in Italian, and it was chosen to commemorate the imminent introduction of Arduino 1.0. Moving forward, the Uno and version 1.0 will be the reference versions of Arduino. The Uno is the latest in a series of USB Arduino boards and the standard model for the Arduino platform; see the index of Arduino boards for a comparison with prior versions.

**POWER:**

VIN is the input voltage to the Arduino board when it is powered by an external source (rather than the 5 volts from the USB connection or another regulated power source). This pin can be used to supply voltage or to access voltage if it is supplied via the power jack.

• 5V. This pin outputs a controlled 5V from the board's regulator. The board can be powered by the DC power jack (7 - 12V), the USB connector (5V), or the board's VIN pin (7-12V). Using the 5V or 3.3V pins to supply voltage bypasses the regulator and can damage your board. We do not recommend it.

• 3V 3. The on-board regulator generates a 3.3-volt supply. The maximum current draw is 50 milliamperes.

• GND. Pins that are grounded.

VIN is the input voltage to the Arduino board when it is powered by an external source (rather than the 5 volts from the USB connection or another regulated power source). This pin can be used to supply voltage or to access voltage if it is supplied via the power jack.

• 5V. This pin outputs a controlled 5V from the board's regulator. The board can be powered by the DC power jack (7 - 12V), the USB connector (5V), or the board's VIN pin (7-12V). Using the 5V or 3.3V pins to supply voltage bypasses the regulator and can damage your board. We do not recommend it.

• 3V 3. The on-board regulator generates a 3.3-volt supply. The maximum current draw is 50 milliamperes.

• GND. Pins that are grounded.

**ARDUINO UNO:**

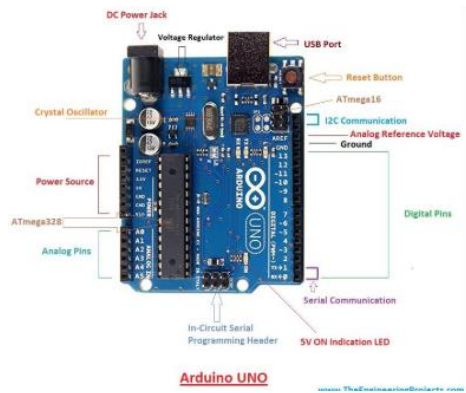

**Memory**

The ATmega328 has 32 KB (0.5 KB is reserved for the bootloader). It also features 2 KB of SRAM and 1 KB of EEPROM (read and writeable using the EEPROM library). Output and Input The Uno's 14 digital pins can be used as inputs or outputs by using the pin Mode (), digital Write (), and digital Read () routines. They run on 5 volts. Each pin includes an inbuilt pull-up resistor (disconnected by default) of 20-50 kilo Ohms and may give or receive a maximum of 40 mA.

• SS: 10, MOSI: 11, MISO: 12, SCK: 13. The SPI library is used to communicate with these pins via SPI.

• LED: 13. A built-in LED is connected to digital pin 13. When the pin is HIGH, the LED is turned on; when it is LOW, the LED is turned off. The Uno features six analogue inputs labelled A0 through A5, each with a resolution of 10 bits (i.e., 1024 distinct values). They measure from ground to 5 volts by default, but the upper end of their range can be changed using the AREF pin and the analogue Reference () function. Furthermore, several pins have specialised functions:

• TWI: A4 (SDA) and A5 (SCL) pins. The Wire library is used to support TWI communication. There are a few more pins on the board:

• AREF. The analogue inputs' reference voltage. Used in conjunction with analogue Reference ().

• Reset. To reset the microcontroller, connect this wire to ground. Typically used to provide a reset button to shields that block the board's. Also also the mapping of Arduino pins to ATmega328 ports. The mapping is the same for the Atmega8, 168, and 328. Communication The Arduino Uno includes a number of communication ports for connecting to a computer, another Arduino, or other microcontrollers.

The ATmega328 supports UART TTL (5V) serial communication via digital pins 0 (RX) and 1 (TX). An ATmega16U2 on the board channels serial communication over USB and appears to software on the PC as a virtual com port. The '16U2 firmware makes use of normal USB COM drivers, so no additional drivers are required.

On Windows, however, a.inf file is necessary. The Arduino software features a serial monitor for sending and receiving simple textual data to and from the Arduino hardware. When data is transmitted via the USB-to-serial chip and USB connection to the computer, the RX and TX LEDs on the board will flash (but not for serial communication on pins 0 and 1). A piece of software The serial library enables serial communication on any digital pin of the Uno. I2C (TWI) and SPI communication are also supported by the ATmega328. The Arduino software contains a Wire library that simplifies I2C bus use; check the documentation for further information. Use the SPI library to communicate using SPI.

*H. power supply*
A power supply (sometimes spelt power supply unit or PSU) is a device or system that provides electrical or other types of energy to an output load or group of loads. The word is most typically applied to electrical energy suppliers, less frequently to mechanical energy providers, and rarely to others.

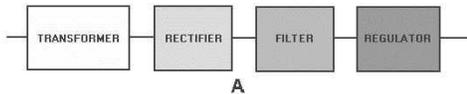

**Block diagram of a basic power supply**

The transformer increases or decreases the voltage on the input line and isolates the power source from the power line. The rectifier section is responsible for converting the alternating current input signal to pulsing direct current. However, as you read more in this chapter, you will discover that pulsing dc is not ideal. As a result, a filter section is utilised to convert pulsing direct current to a purer, more acceptable type of direct current voltage.

    The third element, the regulator, accomplishes exactly what its name suggests. It keeps the power supply's output steady in the face of substantial variations in load current or input line voltages. Let's trace an AC signal through the power supply now that you know what each part accomplishes. You must now examine how this signal is altered within each section of the power supply. Later in the chapter, you'll see how these modifications occur. In Figure 4-1, view B, an input signal of 115 volts AC is applied to the transformer's primary. The transformer is a step-up transformer with a one-to-three turn ratio. We may compute the output voltage for this transformer by multiplying the input voltage by the ratio of primary turns to secondary turns; so, 115 volts AC' 3 = 345 volts ac (peak-to-peak) at the output. Because each diode in the rectifier section conducts for 180 degrees of the 360-degree input, the rectifier's output will be one-half of the 360-degree input, or roughly 173 volts of pulsing DC. The rise and fall time of the changing signal is controlled by the filter section, which is a network of resistors, capacitors, or inductors. As a result, the signal maintains a more steady DC level. In the discussion of the real filter circuits, we shall see the filter process more clearly. The filter's output is a signal of 110 volts dc with ac ripple riding on the dc. The cause of the decreased voltage (average voltage) will be discussed. The regulator keeps its output at a consistent 110-volt direct current level, which is used by the electronic equipment (also known as the load).
Simple 5 volt power supply for digital circuits

• Brief functioning description: Provides a well-regulated +5V output with a current capability of 100 mA.

• Overheating protection: When the regulator IC gets too hot, the output is cut down.

• Circuit complexity: Very basic and straightforward to construct.

• Circuit performance: Very consistent +5V output voltage, long-term operation.

• Component availability: Simple to obtain; employs only extremely common fundamental components.

• Applications: Electronic device component, tiny laboratory power supply.

• Power supply voltage: DC 8-18V unregulated power supply.

• Power supply current: required output current plus 5 mA.

DESCRIPTION OF THE CIRCUIT

This circuit is a modest +5V power source that comes in handy when exploring with digital electronics. These transformers are widely accessible, but their voltage control is typically poor, making them unsuitable for digital circuit experimentation unless greater regulation can be provided in some way.

This circuit can provide +5V output at approximately 150 mA current, but it can be boosted to 1 A when the 7805 regulator chip is properly cooled. Overload and terminal protection are built into the circuit.

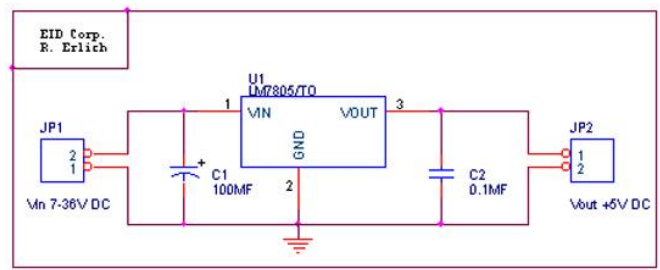

**Circuit diagram**

**Circuit diagram of the power supply:**

The capacitors must be rated at a high enough voltage to securely withstand the input voltage provided to the circuit. The circuit is fairly simple to incorporate onto a Vero board.

**COMPONENT LIST**

• 7805 regulator integrated circuit.

• A 100 uF electrolytic capacitor with a voltage rating of at least 25V.

• A 10 uF electrolytic capacitor with a voltage rating of at least 6V.

• A ceramic or polyester capacitor with a capacity of 100 nF.

MORE CURRENT OUTPUT

If we require more than 150 mA of output current, we can increase the output current to 1A by making the following changes:

• Replace the transformer that supplies power to the circuit with one that can provide as much current as we require from the output.

• Attach a heat sink to the 7805 regulators (large enough so that it does not overheat due to the increased losses in the regulator).

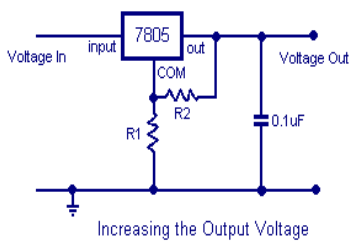

*VOLTAGES OF OTHER OUTPUT*

if we require voltages other than +5v, we can adapt the circuit by replacing the 7805 chips with another regulator from the 78xx chip family with a different output voltage. the output voltage is indicated by the last numerals in the chip code. the input voltage must be at least 3v higher than the regulator output voltage for the regulator to function properly.

Software implementation:

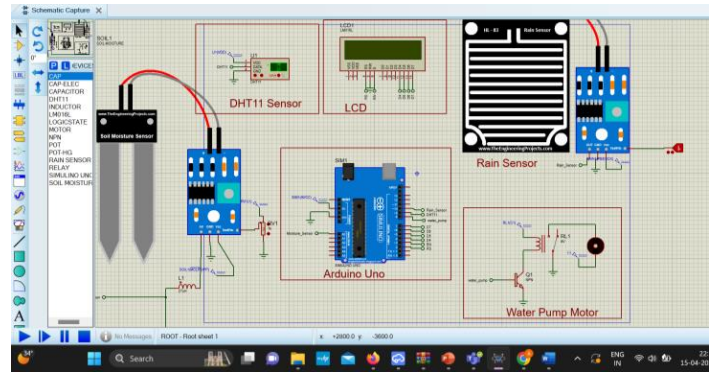

Hardware implementation:

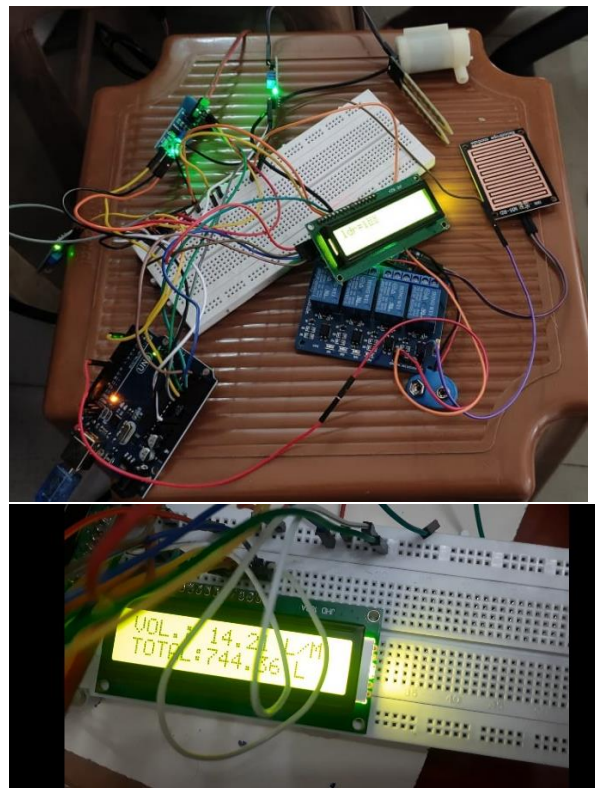

V. CONCLUSION

The Sensor based agricultural monitoring system is an innovative support system. The automated system implements are reliable and minimize the dependence on human elements for monitoring the various agricultural parameters. The cost of implementing the devices can also be minimized by integrating the various units in a zone. The information given by the devices could be properly analyzed and the usefulness of the system can be enhanced. The use of the new technologies coupled with technical excellence will indisputably lea to higher productivity and to an enviable position in the realm of agricultural excellence.